\documentclass[runningheads]{llncs}
\pdfoutput=1
\usepackage{graphicx}
\usepackage{amsmath}
\usepackage{multirow}
\usepackage{booktabs}
\usepackage{float}
\usepackage{mwe}
\usepackage{subfig}
\begin{document}
\title{Can we trust deep learning based diagnosis? \\ 
The impact of domain shift in chest radiograph classification}
\titlerunning{The impact of domain shift in chest radiograph classification}
%
\author{Eduardo H P Pooch\thanks{Corresponding author}, Pedro L. Ballester, Rodrigo C. Barros}

\institute{School of Technology \\ Pontificia Universidade Católica do Rio Grande do Sul \\
\email{eduardo.pooch@edu.pucrs.br}}

\authorrunning{Pooch et al.}

%
\maketitle              

\begin{abstract}
While deep learning models become more widespread, their ability to handle unseen data and generalize for any scenario is yet to be challenged. In medical imaging, there is a high heterogeneity of distributions among images based on the equipment that generates them and their parametrization. This heterogeneity triggers a common issue in machine learning called domain shift, which represents the difference between the training data distribution and the distribution of where a model is employed. A high domain shift tends to implicate in a poor generalization performance from the models. In this work, we evaluate the extent of domain shift on four of the largest datasets of chest radiographs. We show how training and testing with different datasets (e.g., training in ChestX-ray14 and testing in CheXpert) drastically affects model performance, posing a big question over the reliability of deep learning models trained on public datasets. We also show that models trained on CheXpert and MIMIC-CXR generalize better to other datasets.

\end{abstract}

\section{Introduction}
\label{sec:intro}
Radiography is a common exam to diagnose chest conditions since it is a low-cost, fast, and widely available imaging modality. 
Abnormalities identified on radiographs are called radiological findings. Several chest radiological findings might indicate lung cancer, such as lesions, consolidation, and atelectasis. Lung cancer is the first cause of cancer death worldwide, and the lack of effective early-detection methods is one of the main reasons for its poor prognosis~\cite{hirsch2001early}. Lung cancer signs are mostly identified through imaging exams, but at the same time, 90\% of the lung cancer misdiagnosis occurs in radiographs, often due to observer error~\cite{Ciello2017}.

Deep learning is a growing field for image analysis. It has recently been employed at several medical imaging tasks~\cite{Litjens2017} and may help to overcome observer error. Considering chest radiographs, deep learning approaches are usually developed within a multi-label classification scenario, providing findings to assist physicians with the diagnosis process. Recent work in the field achieved near radiologist-level accuracy at identifying radiological findings by the use of convolutional neural networks~\cite{rajpurkar2017chexnet}, one of the most successful deep learning approaches.

\begin{figure}
\includegraphics[width=\textwidth]{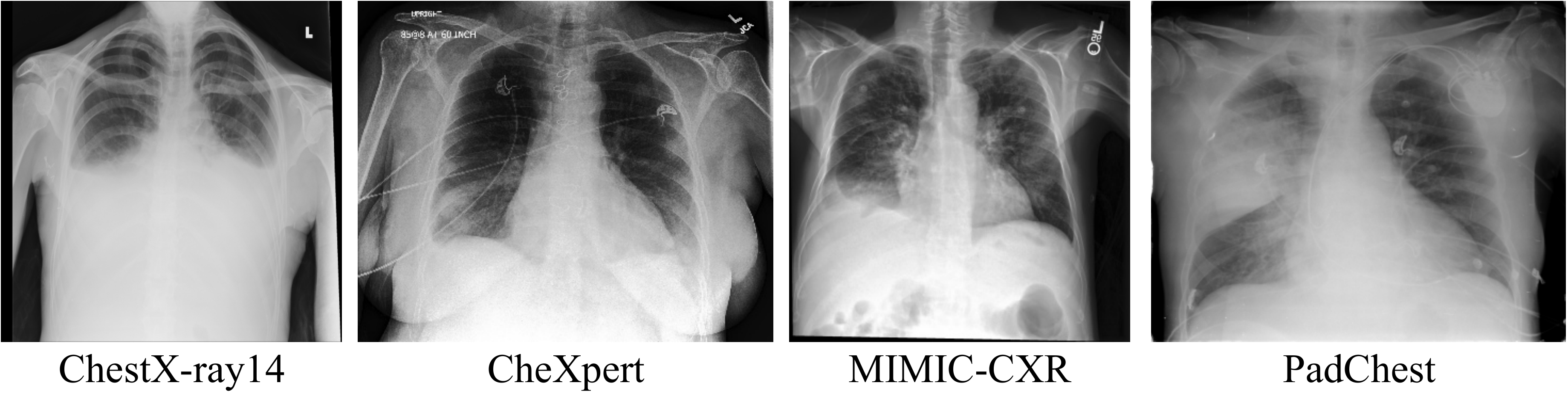}
\caption{Example of a chest radiograph positive for consolidation randomly sampled from each of the four analyzed datasets: ChestX-ray14, CheXpert, MIMIC-CXR, and PadChest.}
\label{fig}
\end{figure}

One assumption underlying deep learning models is that training and test data are independent and identically distributed (\textit{i.i.d}). This assumption often does not hold when data comes from different settings.
This is a common case for medical imaging, a scenario in which image acquisition protocols and machines may vary among diagnostic centers, being defined by the quality of the machine, its parameters, and the acquisition protocol. 
Another aspect of medical imaging is the epidemiological variation among different populations, which may change the label distribution in different datasets. 
This difference in data distribution from the same task is called \textit{domain shift}. The domain from where training data is sampled is the source domain, with distribution $p(X_s)$, and the one where the model is applied to is the target domain, with distribution $p(X_t)$. When $p(X_s) \sim p(X_t)$, it means the model most likely will handle test data the same way as it did in training. As $p(X_s)$ diverges from $p(X_t)$, trained models tend to yield poor results, failing to effectively handle the input data~\cite{torralba2011unbiased}. 

With deep learning becoming widespread, predictive models will inevitably become a big part of health care. We believe that before health care providers trust predictive models as a second opinion, we must understand the extent of their generalization capabilities and how well they perform outside the source domain. In this work, we propose to evaluate how well models trained on a hospital-scale database generalize to unseen data from other hospitals or diagnostic centers by analyzing the degree of domain shift among four large datasets of chest radiographs. We train a state-of-the-art convolutional neural network for multi-label classification at each of the four datasets and evaluate their performance at predicting labels at the remaining three.

This paper is organized as follows: we first describe our methods, detailing our experiment design and datasets. Then, we summarize our results in Table~\ref{tab:full_results} and discuss our findings.

\section{Related work}
The impact of domain shift in medical imaging has been studied for brain tumors by AlBadawy~et~al.~\cite{albadawy2018deep}. They showed how training models with data from a different institution to where it is tested impacted the results for brain tumor segmentation. They also found that using multiple institutions for training does not necessarily remove this limitation.

We also see methods focused on unsupervised domain adaptation, where the task is to mitigate the problems of domain shift with unlabeled data from the target domain. Madani~et~al.~\cite{Madani2018} observed the problem of domain overfitting on chest radiographs and developed a semi-supervised learning method based on generative adversarial networks (GANs) capable of detecting cardiac abnormality alongside the adversarial objective. Their method was able to overcome the domain shift between two datasets and increased the model performance when testing on a different domain.
Chen~et~al.~\cite{Chen2018} developed a domain adaptation method based on CycleGANs. Their work resembles CyCADA~\cite{hoffman2017cycada},
with the difference of also introducing a discriminator for the network output, creating what they called semantic-aware GANs. Javanmardi~and~Tasdizen~\cite{javanmardi2018domain} used a framework very similar to domain-adversarial neural networks~\cite{ganin2016domain} that employ a domain classification network with a gradient reversal layer to model a joint feature space. 

Some work focus on domain adaptation for cross-modality segmentation. Dou et al.~\cite{dou2018unsupervised} designed an adversarial-based method that learns both domain-specific feature extraction layers and a joint high-level representation space that can segment data from both MRI or CT data.

Gholami~et~al.~\cite{gholami2018novel} propose a biophysics-based data augmentation method to produce synthetic tumor-bearing images to the training set. The authors argue that this augmentation procedure improves model generalization. Mahmood~et~al.~\cite{Mahmood2018} present a generative model that translates images from the simulated endoscopic images domain to a realistic-looking domain as data augmentation. The authors also introduce an L1-regularization loss between the translated and the original image to minimize distortions.

\section{Materials and methods}
\label{sec:methods}

We wish to understand how trained models at each dataset behave when dealing with data from others.

\subsection{Datasets}
Three large datasets of chest radiographs are available to this date. ChestX-ray14~\cite{wang2017chestxray8} from the National Institute of Health contains 112,120 frontal-view chest radiographs from 32,717 different patients labeled with 14 radiological findings. CheXpert \cite{irvin2019chexpert} from the Stanford Hospital contains 224,316 frontal and lateral chest radiographs of 65,240 patients. MIMIC-CXR \cite{johnson2019mimic} from Massachusetts Institute of Technology presents 371,920 chest X-rays associated with 227,943 imaging studies from 65,079 patients. Both CheXpert and MIMIC-CXR are labeled with the same 13 findings. PadChest \cite{bustos2019padchest} 160,000 images obtained from 67,000 patients of San Juan Hospital in Spain. The radiographs are labeled with 174 different findings. 
Most labels from all four datasets are automatically extracted using natural language processing algorithms on the radiological reports.

\begin{figure}[ht]
    \centering
    \includegraphics[width=\textwidth]{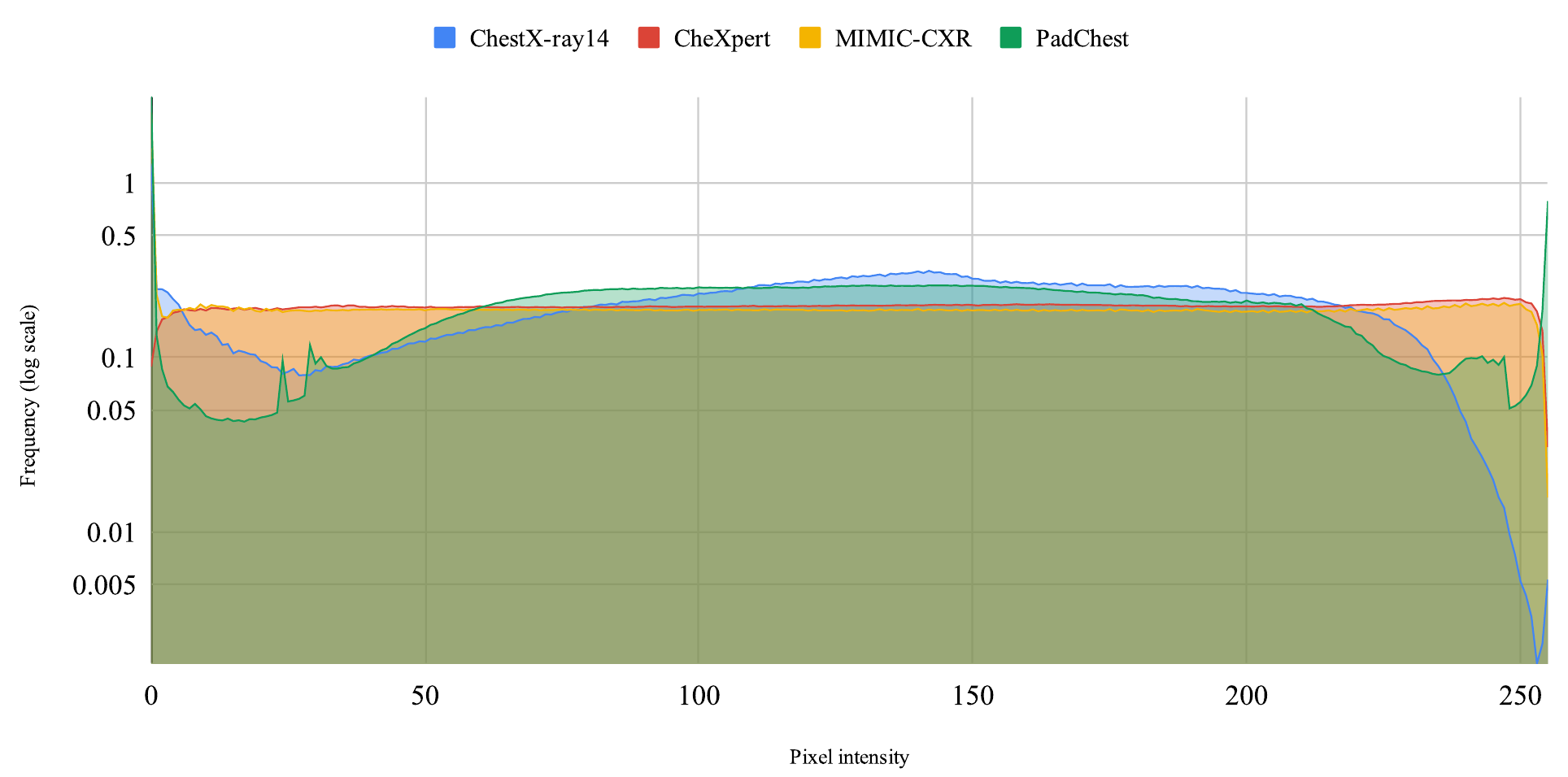}
    \caption{Dataset pixel intensity probability density function.}
    \label{fig:distribution}
\end{figure}

\begin{figure}[ht]
    \centering
    \includegraphics[width=\textwidth]{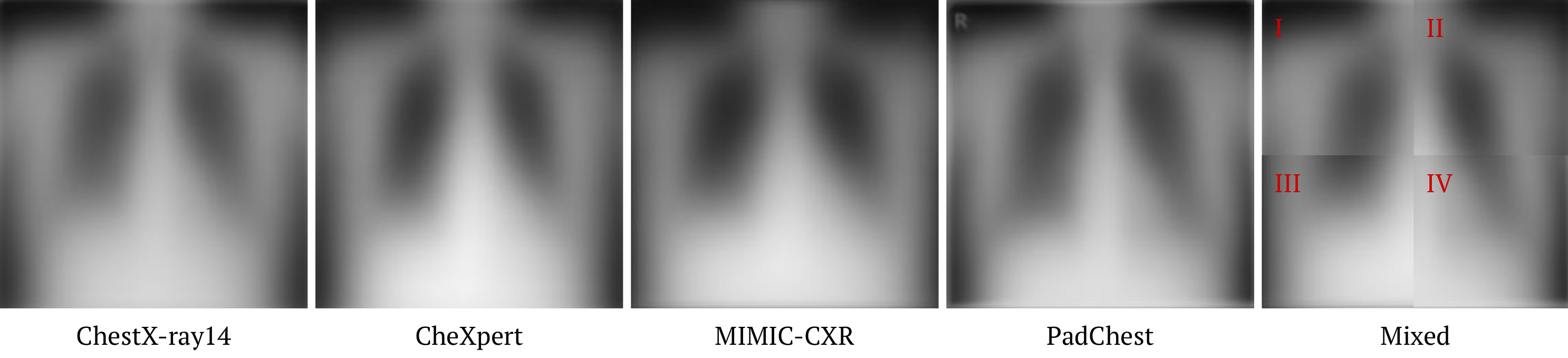}
    \caption{Average image of each of the four datasets. Last image cotains one quarter of each average image to better visualize pixel intensity differences (I - ChestX-ray14, II - CheXpert, III - MIMIC-CXR, IV - PadChest).}
    \label{fig:average}
\end{figure}

We show the pixel intensity distribution of each dataset in Figure~\ref{fig:distribution}. We see a spike at low intensities (especially $0$) for every center. However, the distribution for higher intensities is somewhat different for every center, which might implicate in a decrease of the models' predictive performance, except for CheXpert and MIMIC-CXR, which show similar distributions.
Figure \ref{fig:average} shows the average radiograph of each dataset (computed using 10,000 random samples), in which we can see small differences in pixel intensity and that a common artifact appears on the top left corner of PadChest radiographs.
Another difference that might implicate in domain shift is that PadChest labels are extracted from reports in Spanish, while the other three are extracted from reports in English.

\subsection{Experiment design}

We employ a multi-label classification approach reproducing CheXNet \cite{rajpurkar2017chexnet}, which achieved state-of-the-art results in classification of multiple pathologies using a DenseNet121 convolutional neural network architecture \cite{huang2017densely}.
The model is pre-trained on the ImageNet dataset, and the images are resized to $224\times 224$ pixels and normalized using ImageNet mean and standard deviation. We train four models, one for each dataset, and subsequently evaluate our model at the other three. Each model is trained with the training set and evaluated in its own test set and the other three test sets. The four datasets have the same train, test, and validation sets across the experiments. For the ChestX-ray14 dataset, we use the original split, but since CheXpert and MIMIC-CXR test sets are not publicly available and PadChest does not have an original split, we randomly re-split their data, keeping ChestX-ray14 split ratio (70\% train, 20\% test, and 10\% validation) and no patient overlap between the sets. Table~\ref{tab:label_freq} shows the frequency of the labels in each training and test split. As both CheXpert and MIMIC-CXR have labels for uncertainty, we assumed these labels as negatives (U-Zeros approach).

\begin{table*}[ht]
\centering
\caption{Positive label frequency (in number of radiographs) in our train and test split for each dataset.}
\label{tab:label_freq}
\resizebox{\textwidth}{!}{%
\begin{tabular}{@{}ccccccccccccccccc@{}}
\toprule
 & \multicolumn{2}{c}{Atelectasis} & \multicolumn{2}{c}{Cardiomegaly} & \multicolumn{2}{c}{Consolidation} & \multicolumn{2}{c}{Edema} & \multicolumn{2}{c}{Lesion} & \multicolumn{2}{c}{Pneumonia} & \multicolumn{2}{c}{Pneumothorax} & \multicolumn{2}{c}{No Finding} \\ \midrule
 & Train & Test & Train & Test & Train & Test & Train & Test & Train & Test & Train & Test & Train & Test & Train & Test \\ \midrule
ChestX-ray14 & 7,996 & 2,420 & 1,950 & 582 & 3,263 & 957 & 1,690 & 413 & 7,758 & 2,280 & 978 & 242 & 3,705 & 1,089 & 42,405 & 11,928 \\
CheXpert & 20,630 & 6,132 & 15,885 & 5,044 & 9,063 & 2,713 & 34,066 & 10,501 & 4,976 & 1,411 & 3,274 & 935 & 12,583 & 3,476 & 12,010 & 3,293 \\
MIMIC-CXR & 34,653 & 10,071 & 34,097 & 9,879 & 8,097 & 2,430 & 20,499 & 5,954 & 5,025 & 1,341 & 12,736 & 3,711 & 8,243 & 2,231 & 58,135 & 16,670 \\ 

PadChest & 1,841 & 574 & 3283 & 953 & 664 & 210 & 127 & 44 & 878 & 261 & 678 & 194 & 163 & 33 & 25,268 & 7,200 \\
\bottomrule
\end{tabular}
}
\end{table*}

One limitation we encountered is that the datasets have a different set of labels between each other. We fix this by training each model with all labels available, but reporting the results only on the common labels for all four ( Atelectasis, Cardiomegaly, Consolidation, Edema, Lesion, Pneumonia, Pneumothorax, and No Finding). We create a ``Lesion'' label on ChestX-ray14 by joining the samples annotated as ``Nodule'' or ``Mass''. For PadChest, we joined labels that can fit into the 8 common findings, (i.e. ``Atelectasis Basal'', ``Total Atelectasis'',  ``Lobar Atelectasis'', and ``Round Atelectasis'' were merged into ``Atelectasis''). Another limitation is that ChestX-ray14 has only frontal X-rays. Therefore, we only use the frontal samples from the other three datasets, which means 191,229 samples on CheXpert, 249,995 on MIMIC-CXR, and 111,176 on PadChest.

To evaluate domain shift, we use a standard performance metric in multi-label classification, the Area Under the Receiver Operating Characteristic curve (AUC), to report both individual radiological findings results and their average for an overall view of model performance. Both the true positive rate and the false positive rate are considered for computing the AUC. Higher AUC values indicate better performance.

\section{Results}
\label{sec:results}
We train the same neural network architecture with the same hyperparameters at each of the four datasets individually. Training and testing on ChestX-ray14 achieve results similar to the ones reported by CheXnet \cite{rajpurkar2017chexnet}, which exceeded radiologists' performance in detecting pneumonia. After training, we load our model and evaluate it with images from the remaining two.

\begin{table*}[htbt]
    \scriptsize
    \centering
    \caption{Resulting AUCs for the 8 radiological findings common to the four datasets. Best results for each test set are in bold.}
    \label{tab:full_results}
    \resizebox{\textwidth}{!}{
    \begin{tabular}{cccccccccc|c}
        \toprule
        Test set & 
        Training set &
        Atelectasis &
        Cardiomegaly &
        Consolidation &
        Edema & 
        Lesion &
        Pneumonia &
        Pneumothorax &
        No Finding & 
        Mean\\ 
       
        \midrule
        
        \multirow{4}{*}{ChestX-ray14} & 
        \textbf{ChestX-ray14} &
        \textbf{0.8205}	& 
        \textbf{0.9104}	& 
        \textbf{0.8026}	& 
        \textbf{0.8935}	& 
        \textbf{0.7819}	& 
        0.7567	& 
        \textbf{0.8746}	& 
        \textbf{0.7842}	& 
        \textbf{0.8343}	
        \\ 
      
        & CheXpert & 
        0.7850	& 
        0.8646	& 
        0.7771	& 
        0.8584	& 
        0.7291	& 
        0.7287	& 
        0.8464	& 
        0.7569	& 
        0.7933	
        \\ 
        
        & MIMIC-CXR & 
        0.8024	& 
        0.8322	& 
        0.7898	& 
        0.8609	& 
        0.7457	& 
        \textbf{0.7656}	& 
        0.8429	& 
        0.7652	& 
        0.8006	
        \\
        
        & PadChest &
        0.7371	& 
        0.8124	& 
        0.7031	& 
        0.8213	& 
        0.6301	& 
        0.6487	& 
        0.7417	& 
        0.7384	& 
        0.7291	
        \\
        \midrule
        
        \multirow{4}{*}{CheXpert} &
 
        ChestX-ray14 & 
        0.6433	& 
        0.7596	& 
        0.6431	& 
        0.7145	& 
        0.6821	& 
        0.5967	& 
        0.7356	& 
        0.7717	& 
        0.6821	
        \\ 
        
        & \textbf{CheXpert} &
        \textbf{0.6930}	& 
        \textbf{0.8687}	& 
        \textbf{0.7323}	& 
        \textbf{0.8344}	& 
        \textbf{0.7882}	& 
        \textbf{0.7619}	& 
        \textbf{0.8709}	& 
        \textbf{0.8842}	& 
        \textbf{0.8042}	
        \\ 
        
        & MIMIC-CXR & 
        0.6576	& 
        0.8197	& 
        0.7002	& 
        0.7946	& 
        0.7465	& 
        0.7219	& 
        0.8046	& 
        0.8564	& 
        0.7627	
        \\ 
        & PadChest &
        0.6127	& 
        0.7397	& 
        0.6352	& 
        0.6934	& 
        0.6978	& 
        0.6510	& 
        0.6209	& 
        0.7600	& 
        0.6764	
        \\
        \midrule
        
        \multirow{4}{*}{MIMIC-CXR} & 
        ChestX-ray14 &
       0.7616	& 
        0.7230	& 
        0.7567	& 
        0.8146	& 
        0.6880	& 
        0.6630	& 
        0.7773	& 
        0.8106	& 
        0.7406	
        \\ 
        
        & CheXpert &
        0.7587	& 
        0.7650	& 
        0.7936	& 
        0.8685	& 
        0.7527	& 
        0.6913	& 
        0.8142	& 
        0.8452	& 
        0.7861	
        \\ 
        
        & \textbf{MIMIC-CXR} &
        \textbf{0.8177}	& 
        \textbf{0.8126}	& 
        \textbf{0.8229}	& 
        \textbf{0.8922}	& 
        \textbf{0.7788}	& 
        \textbf{0.7461}	& 
        \textbf{0.8845}	& 
        \textbf{0.8718}	& 
        \textbf{0.8283}	
        \\
        & PadChest &
        0.7218	& 
        0.6899	& 
        0.7200	& 
        0.7828	& 
        0.6577	& 
        0.6454	& 
        0.6995	& 
        0.7976	& 
        0.7143	
        \\
        
         \midrule
        
        \multirow{4}{*}{PadChest} & 
        ChestX-ray14 &
        0.7938	& 
        0.8822	& 
        0.8300	& 
        0.8893	& 
        \textbf{0.7010}	& 
        0.7366	& 
        0.7176	& 
        0.8028	& 
        0.7929	
        \\ 
        
        & CheXpert &
        0.7566	& 
        0.8656	& 
        0.8511	& 
        \textbf{0.9390}	& 
        0.6833	& 
        0.7269	& 
        \textbf{0.8731}	& 
        0.8335	& 
        0.8161	
        \\ 
        
        & MIMIC-CXR &
        \textbf{0.7942}	& 
        0.8270	& 
        \textbf{0.8963}	& 
        0.9310	& 
        0.6761	& 
        \textbf{0.8060}	& 
        0.8308	& 
        0.8217	& 
        0.8229	
        \\
        & \textbf{PadChest} &
        0.7641	& 
        \textbf{ 0.9075}	& 
        0.8607	& 
        0.9107	& 
        0.6975	& 
        0.7990	& 
        0.8276	& 
        \textbf{0.8710}	& 
        \textbf{0.8298}	
        \\
        \bottomrule
    \end{tabular}
    }
\end{table*}

We summarize the results in Table~\ref{tab:full_results}. We can see that the best average result for each test set appears when the training set is from the same dataset. This shows that clinicians should expect a decrease in the reported performances of machine learning models when applying them in real-world scenarios. The decrease may vary according to the dataset distribution in which the model was trained. For instance, running a model trained on MIMIC-CXR on CheXpert's test set reduces the mean AUC in $0.04$, while the model trained on ChestX-ray14 reduces it in $0.12$. On MIMIC-CXR's test set, training on MIMIC-CXR shows almost the same decrease in mean AUC ($0.04$), reducing the AUC in all of the findings. The model trained on ChestX-ray14 has the highest average AUC when testing on its own test set, but when testing in other datasets, it shows a significant performance drop, lowering CheXpert's mean AUC in $0.12$, MIMIC-CXR's in $0.08$ and PadChest in $0.04$. Both the models trained on CheXpert and MIMIC-CXR mostly preserve the ChestX-ray14 baseline mean AUC, while the model trained on PadChest drops the average performance in $0.10$. PadChest presented some variations on the best AUC for each disease, probably due to the smaller number of train instances. The models trained on CheXpert and MIMIC-CXR got very close results to PadChest's baseline. 

\begin{figure}[ht]
    \subfloat[\label{chart:nih} Trained on ChestX-ray14]{
       \includegraphics[width=.49\textwidth]{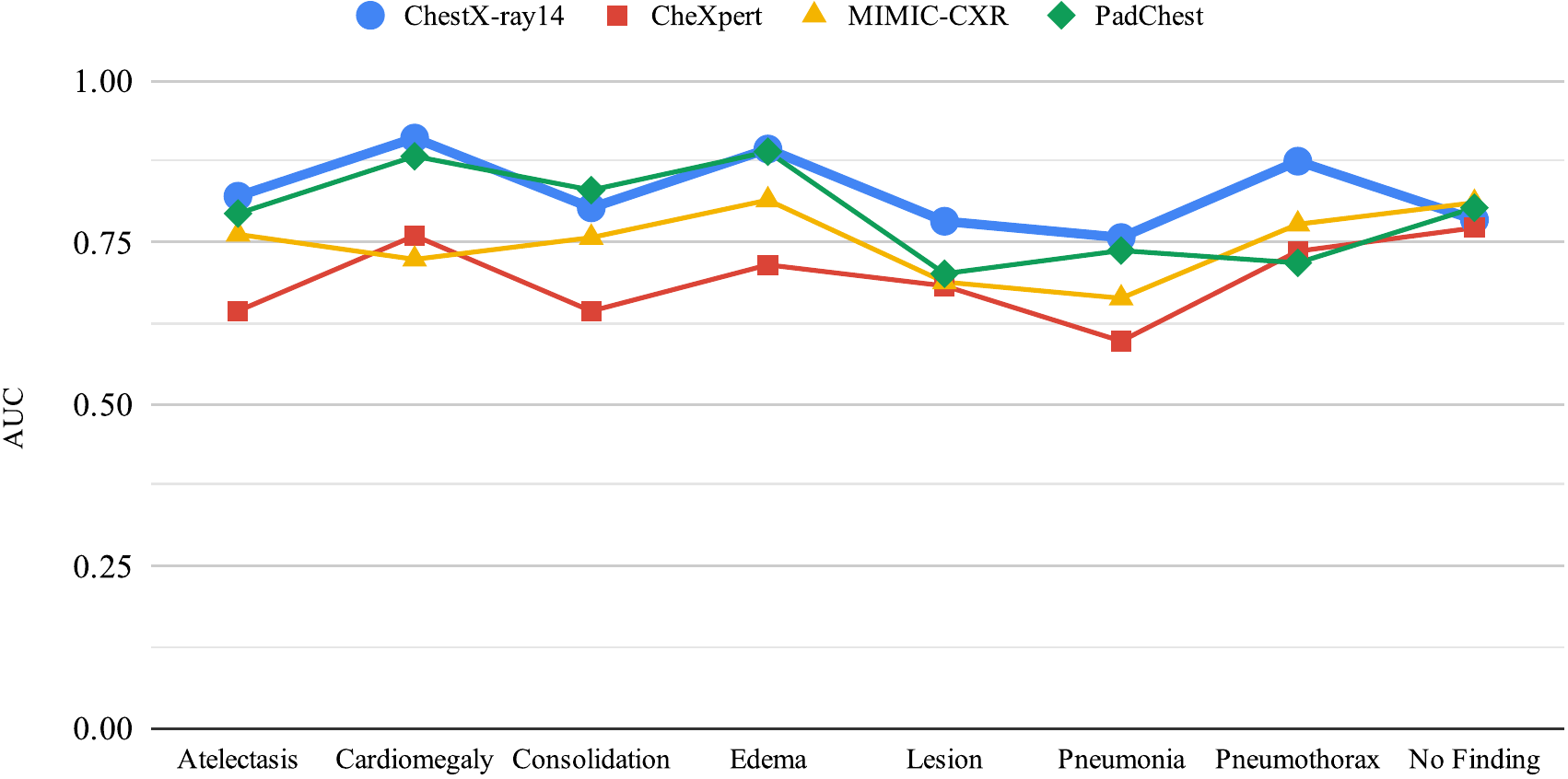}
       }
   \subfloat[\label{chart:chexpert} Trained on CheXpert]{
       \includegraphics[width=.49\textwidth]{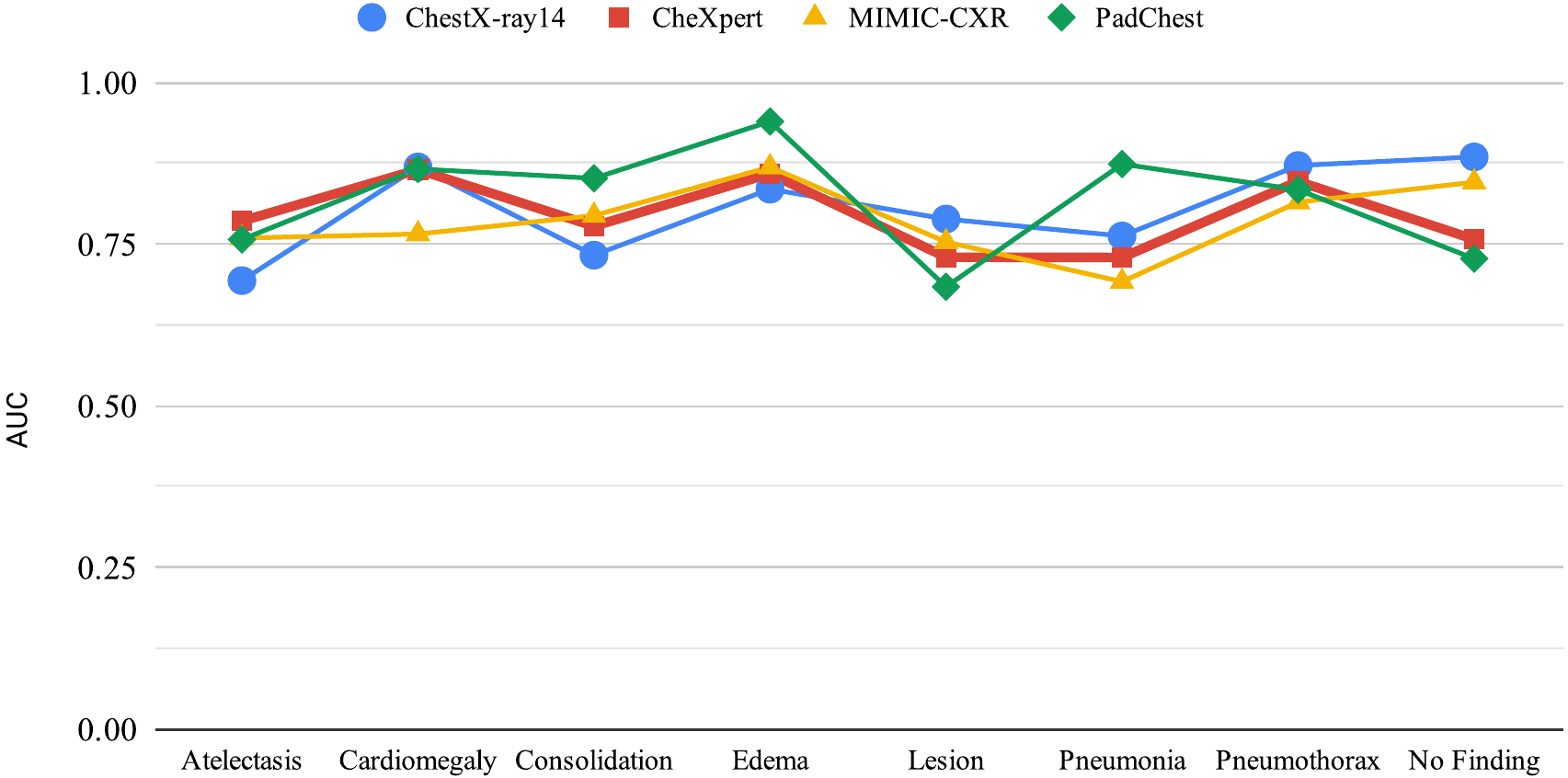}
           }
    \\
    \subfloat[\label{chart:mimic} Trained on MIMIC-CXR]{
       \includegraphics[width=.49\textwidth]{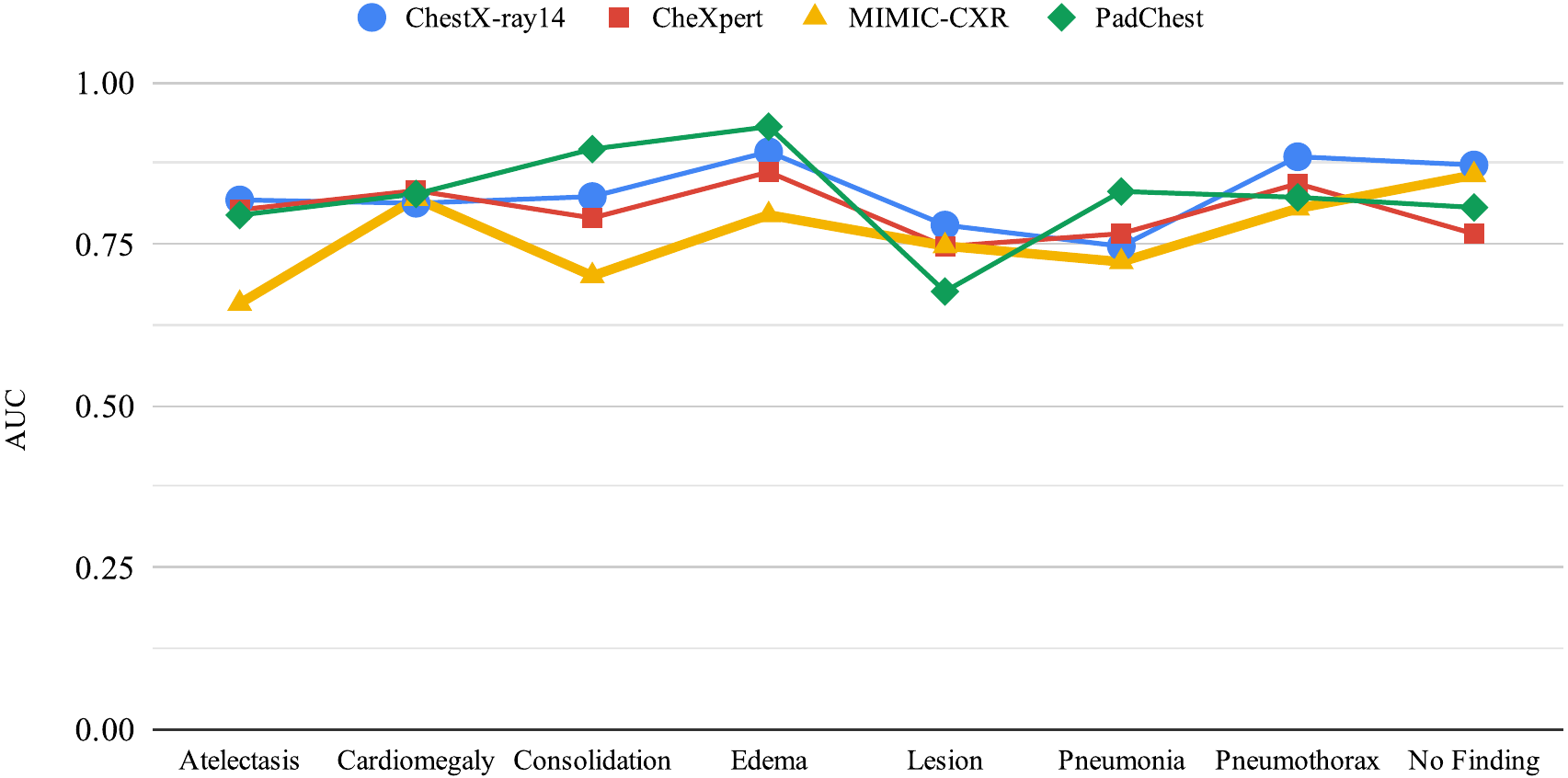}
           }
    \subfloat[\label{chart:padchest} Trained on PadChest]{
       \includegraphics[width=.49\textwidth]{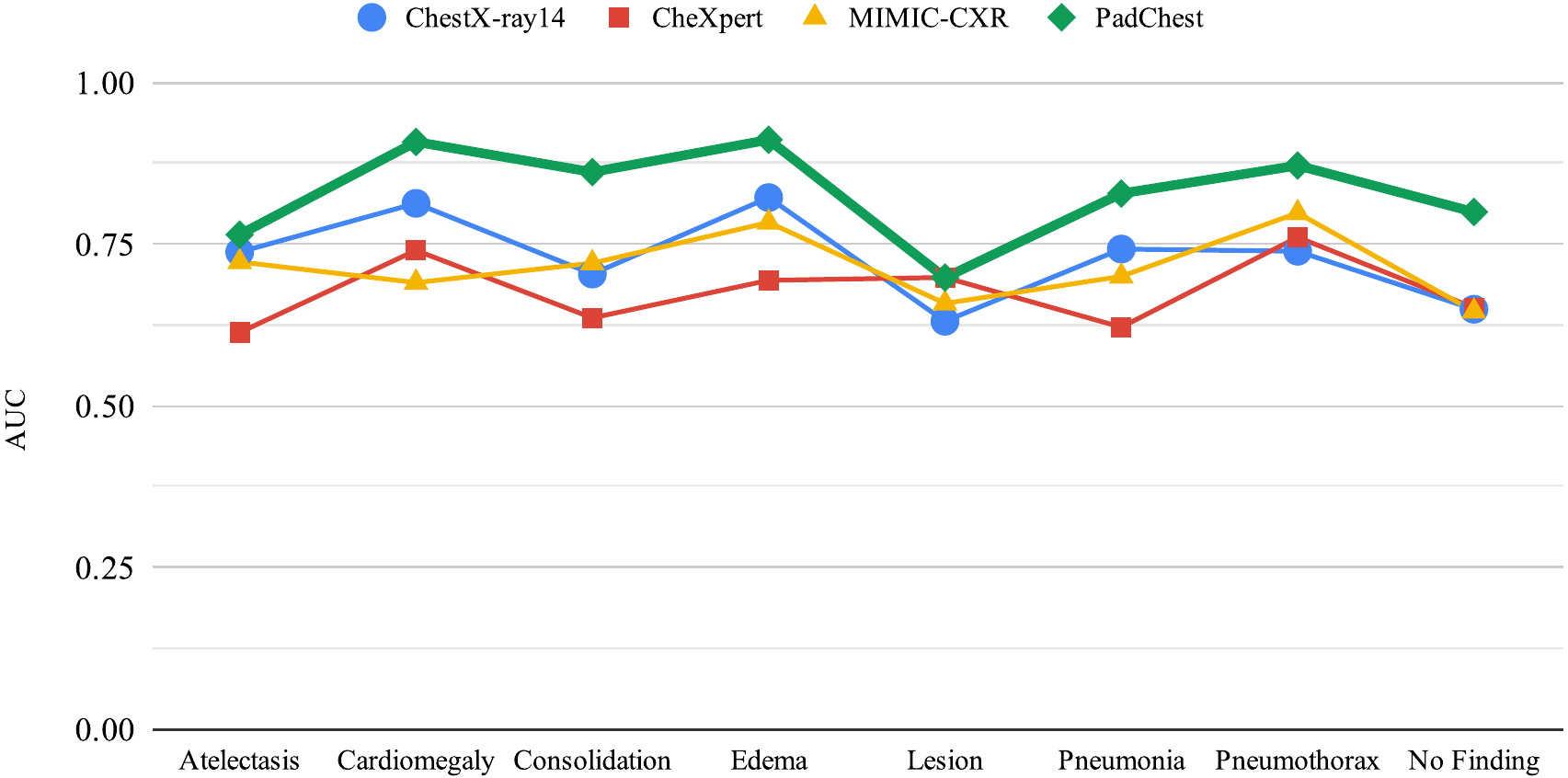}
           }
   \caption{Performance of a model trained on ChestX-ray14 (a), CheXpert (b), MIMIC-CXR (c), and PadChest (d) on each of the four test sets. }
   \label{fig:charts}
\end{figure}

Figure \ref{fig:charts} shows the performance on the test set of the four trained models, represented as lines to better visualize AUC variations. The CheXpert and MIMIC-CXR models show smaller variations on the AUCs of the findings compared to their own test sets, presenting close lines, while PadChest and ChestX-ray14 shows have the line of the own test set mostly on top and a drop in performance on the other test sets. 

Clear evidence of the impact of domain shift over model performance is how frequently the best AUC for each radiological finding comes from the same dataset. In the ChestX-ray14 test set, the best AUC appears $7$ out of $8$ times when training with the same set. The same phenomenon happens on both CheXpert ($8$ out of $8$) and MIMIC-CXR ($8$ out of $8$). Furthermore, in all four test sets, the best average AUC comes from their respective training set. 
One possible cause of domain shift is the label extraction method. CheXpert and MIMIC-CXR used the same labeler, while ChestX-ray14 has its own. 

ChestX-ray14 labeler has raised some questions concerning its reliability. A visual inspection study~\cite{Oakden-Rayner2019} states that its labels do not accurately reflect the content of the images. Estimated label accuracies are $10-30\%$ lower than the values originally reported. It also might be that ChestX-ray14 and PadChest do not have representative training sets since models trained on CheXpert and MIMIC-CXR perform well on ChestX-ray14 and PadChest test sets, but the models trained on ChestX-ray14 and PadChest do not perform well on CheXpert and MIMIC-CXR's test sets.

\section{Discussion and Conclusion}
In this work, we show how a state-of-the-art deep learning model fails to generalize to unseen datasets when they follow a somewhat different distribution. Our experiments show that a model with reported radiologist-level performance had a huge drop in performance outside its source dataset, pointing the existence of domain shift in chest X-rays datasets. Despite recent efforts for the creation of large radiograph datasets in the hope of training generalizable models, it seems that the data acquisition methodology of some available datasets does not capture the required heterogeneity for this purpose. 

Among the analyzed datasets, CheXpert and MIMIC-CXR seem to be most representative of the other datasets, as the models trained on them show a smaller performance drop when comparing to the baseline. Therefore, these two datasets should be preferred by researchers when developing models for chest radiograph analysis. The least representative seems to be ChestX-ray14, whose model did not perform as well outside its own test set, while the models trained on the other datasets performed well when testing on ChestX-ray14. Models trained on PadChest also show a significant performance drop in other test sets, but it might be because of the smaller amount of available data for each finding, as it has a way higher number of annotated findings than the other datasets.

Although deep learning advances allow for new application scenarios, more steps for model validation must be conducted with more emphasis on external validation. We argue that a case-by-case validation is ideal, where the model is validated at new data from each center. The reason is twofold. First, models must be able to handle data from a specific scenario properly. Second, the label distribution from each environment might change due to several external factors, which might not reflect prediction biases learned by the model. One alternative for these limitations is to create small datasets from specific machines where the model will be used, where we fine-tune them after pre-training on large available datasets.

\bibliographystyle{splncs04}
\bibliography{references}

\end{document}